\documentclass[11pt]{article}
\usepackage{amssymb}
\usepackage{amsmath}
\usepackage[english]{babel}
\usepackage{graphicx}

\def\Journal#1#2#3#4{{#1} {\bf #2}, #3 (#4)}

\def\RNC{\em Rivista Nuovo Cimento}

\def\NIMA{{\em Nucl. Instrum. Methods} A}

\def\PLB{{\em Phys. Lett.}  B}
\def\PRL{\em Phys. Rev. Lett.}
\def\PRD{{\em Phys. Rev.} D}

\def\GaC{\em Gravitation and Cosmology}

\def\JETPL{\em JETP Lett.}

\def\CQG{\em Class. Quantum Grav.}
\def\APJ{\em Astrophys. J.}

\def\MPLA{{\em Mod. Phys. Lett.}  A}
\def\IJTP{\em Int. J. Theor. Phys.}
\def\NJP{\em New J. of Phys.}
\def\JHEP{\em JHEP}
\def\EPHJ{\em Eur.Phys.J}

\def\s{{\,\rm s}}

\def\eV{\,{\rm eV}}

\def\TeV{\,{\rm TeV}}

\def\({\left(}
\def\){\right)}
\def\cm{{\,\rm cm}}

\def\beq{\begin{equation}}
\def\eeq{\end{equation}}
\def\bea{\begin{eqnarray}}
\def\eea{\end{eqnarray}}

\begin{document}

    \begin{center}
        \large \textbf{Low energy binding of composite dark matter with nuclei as a solution for the puzzles of dark matter searches}
    \end{center}

    \begin{center}
   Maxim Yu. Khlopov$^{1,2,3}$, Andrey G. Mayorov $^{1}$, Evgeniy Yu.
   Soldatov $^{1}$

    \emph{$^{1}$Moscow Engineering Physics Institute (National Nuclear Research University), 115409 Moscow, Russia \\
    $^{2}$ Centre for Cosmoparticle Physics "Cosmion" 115409 Moscow, Russia \\
$^{3}$ APC laboratory 10, rue Alice Domon et L\'eonie Duquet \\75205
Paris Cedex 13, France}

    \end{center}

\medskip

\begin{abstract}

Positive results of dark matter searches in experiments DAMA/NaI and DAMA/LIBRA taken together with negative results of other groups can imply nontrivial particle physics solutions for cosmological dark matter. Stable particles with charge -2 bind with primordial helium in O-helium "atoms" (OHe), representing a specific
Warmer than Cold nuclear-interacting form of dark matter. Slowed down in the
terrestrial matter, OHe is elusive for direct methods of underground
Dark matter detection like those used in CDMS experiment, but its
low energy binding with nuclei can lead to annual variations of energy
release in the interval of energy 2-6 keV in DAMA/NaI and DAMA/LIBRA
experiments. Schrodinger equation for system of nucleus and OHe is considered and reduced to an equation of relative motion in a spherically symmetrical potential, formed by the Yukawa tail of nuclear scalar isoscalar attraction potential, acting on He beyond the nucleus, and dipole Coulomb repulsion between the nucleus and OHe at distances from the nuclear surface, smaller than the size of OHe. The values of coupling strength and mass of meson, mediating scalar isoscalar nuclear potential, are rather uncertain. Within these uncertainties and in the approximation of rectangular potential wells we find a range of these parameters, at which  the sodium and/or iodine nuclei have a few keV binding energy with OHe. At nuclear parameters, reproducing DAMA results, the energy release predicted for detectors with chemical content other than NaI differ in the most cases from the one in DAMA detector. In particular, it is shown that in the case of CDMS germanium state has binding energy with OHe beyond the range of 2-6 keV and its formation should not lead to ionization in the energy range of DAMA signal.
%The estimated rate of annual modulations of radiative capture of Na by OHe to this level is consistent with the number of events registered in DAMA/NaI and DAMA/LIBRA experiments.
Due to dipole Coulomb barrier, transitions to more energetic levels of Na(I)+OHe system with much higher energy release are suppressed in the correspondence with the results of DAMA experiments. The proposed explanation inevitably leads to prediction of abundance of anomalous Na and I, corresponding to the signal, observed by DAMA.
%We discuss a possibility of similar effects for other chemical content of underground set-ups.

\end{abstract}
\section{Introduction}
The widely shared belief is that the dark matter, corresponding to
$25\%$ of the total cosmological density, is nonbaryonic and
consists of new stable particles. One can formulate the set of
conditions under which new particles can be considered as candidates
to dark matter (see e.g. \cite{book,Cosmoarcheology,Bled07} for
review and reference): they should be stable, saturate the measured
dark matter density and decouple from plasma and radiation at least
before the beginning of matter dominated stage. The easiest way to
satisfy these conditions is to involve neutral weakly interacting
particles. However it is not the only particle physics solution for
the dark matter problem. In the composite dark matter scenarios new
stable particles can have electric charge, but escape experimental
discovery, because they are hidden in atom-like states maintaining
dark matter of the modern Universe.

It offers new solutions for the
physical nature of the cosmological dark matter. The main problem
for these solutions is to suppress the abundance of positively
charged species bound with ordinary electrons, which behave as
anomalous isotopes of hydrogen or helium. This problem is
unresolvable, if the model predicts stable particles with charge -1,
as it is the case for tera-electrons \cite{Glashow,Fargion:2005xz}.
To avoid anomalous isotopes overproduction, stable particles with
charge -1 should be absent, so that stable negatively charged
particles should have charge -2 only.

Elementary particle frames for heavy stable -2 charged species are provided by:
(a) stable "antibaryons" $\bar U \bar U \bar U$ formed by anti-$U$ quark of fourth generation
\cite{I,lom,Khlopov:2006dk,Q} (b) AC-leptons \cite{Khlopov:2006dk,5,FKS}, predicted in the
extension \cite{5} of standard model, based on the approach of
almost-commutative geometry \cite{bookAC}.  (c) Technileptons and anti-technibaryons
\cite{KK} in the framework of walking
technicolor models (WTC) \cite{Sannino:2004qp}. (d) Finally, stable
charged clusters $\bar u_5 \bar u_5 \bar u_5$ of (anti)quarks $\bar
u_5$ of 5th family can follow from the approach, unifying spins and
charges \cite{Norma}.

In the asymmetric case, corresponding to excess of -2 charge
species, $X^{--}$, as it was assumed for $(\bar U \bar U \bar
U)^{--}$ in the model of stable $U$-quark of a 4th generation, as
well as can take place for $(\bar u_5 \bar u_5 \bar u_5)^{--}$ in
the approach \cite{Norma} their positively charged partners
effectively annihilate in the early Universe. Such an asymmetric
case was realized in \cite{KK} in the framework of WTC, where it was
possible to find a relationship between the excess of negatively
charged anti-techni-baryons $(\bar U \bar U )^{--}$ and/or
technileptons $\zeta^{--}$ and the baryon asymmetry of the Universe.
The relationship between baryon asymmetry and excess of -2 charge stable species
is supported by sphaleron transitions at high temperatures and can be realized in all the models,
in which new stable species belong to non-trivial representations of electroweak SU(2) group.

 After it is formed
in the Standard Big Bang Nucleosynthesis (SBBN), $^4He$ screens the
$X^{--}$ charged particles in composite $(^4He^{++}X^{--})$ {\it
O-helium} ``atoms''
 \cite{I}.
 For different models of $X^{--}$ these "atoms" are also
called ANO-helium \cite{lom,Khlopov:2006dk}, Ole-helium
\cite{Khlopov:2006dk,FKS} or techni-O-helium \cite{KK}. We'll call
them all O-helium ($OHe$) in our further discussion, which follows
the guidelines of \cite{I2}.

In all these forms of O-helium $X^{--}$ behave either as leptons or
as specific "heavy quark clusters" with strongly suppressed hadronic
interaction. Therefore O-helium interaction with matter is
determined by nuclear interaction of $He$. These neutral primordial
nuclear interacting objects contribute to the modern dark matter
density and play the role of a nontrivial form of strongly
interacting dark matter \cite{Starkman,McGuire:2001qj}. The active
influence of this type of dark matter on nuclear transformations
seems to be incompatible with the expected dark matter properties.
However, it turns out that the considered scenario of nuclear-interacting O-helium
Warmer than Cold Dark Matter is not easily
ruled out \cite{I,FKS,KK,Khlopov:2008rp} and challenges the
experimental search for various forms of O-helium and its charged
constituents.

Here we
concentrate on its effects in underground detectors. We present
qualitative confirmation of the earlier guess \cite{I2,KK2} that
the positive results of dark matter searches in DAMA/NaI (see for
review \cite{Bernabei:2003za}) and DAMA/LIBRA \cite{Bernabei:2008yi}
experiments can be explained by O-helium, resolving the controversy
between these results and negative results of other experimental
groups.

\section{OHe in the terrestrial matter}
The evident consequence of the O-helium dark matter is its
inevitable presence in the terrestrial matter, which appears opaque
to O-helium and stores all its in-falling flux.

After they fall down terrestrial surface, the in-falling $OHe$
particles are effectively slowed down due to elastic collisions with
matter. Then they drift, sinking down towards the center of the
Earth with velocity \beq V = \frac{g}{n \sigma v} \approx 80 S_3
A^{1/2} \cm/\s. \label{dif}\eeq Here $A \sim 30$ is the average
atomic weight in terrestrial surface matter, $n=2.4 \cdot 10^{24}/A \cm^{-3}$
is the number density of terrestrial atomic nuclei, $\sigma v$ is the rate
of nuclear collisions, $m_o \approx M_X+4m_p=S_3 \TeV$ is the mass of O-helium, $M_X$ is the mass of the $X^{--}$ component of O-helium, $m_p$ is the mass of proton and $g=980~ \cm/\s^2$.

Near the Earth's surface, the O-helium abundance is determined by
the equilibrium between the in-falling and down-drifting fluxes.

The in-falling O-helium flux from dark matter halo is
$$
  F=\frac{n_{0}}{8\pi}\cdot |\overline{V_{h}}+\overline{V_{E}}|,
$$
where $V_{h}$-speed of Solar System (220 km/s), $V_{E}$-speed of
Earth (29.5 km/s) and $n_{0}=3 \cdot 10^{-4} S_3^{-1} \cm^{-3}$ is the
local density of O-helium dark matter. For qualitative estimation we don't take into account here velocity dispersion and distribution of particles in the incoming flux that can lead to significant effect.

At a depth $L$ below the Earth's surface, the drift timescale is
$t_{dr} \sim L/V$, where $V \sim 400 S_3 \cm/\s$ is given by
Eq.~(\ref{dif}). It means that the change of the incoming flux,
caused by the motion of the Earth along its orbit, should lead at
the depth $L \sim 10^5 \cm$ to the corresponding change in the
equilibrium underground concentration of $OHe$ on the timescale
$t_{dr} \approx 2.5 \cdot 10^2 S_3^{-1}\s$.

The equilibrium concentration, which is established in the matter of
underground detectors at this timescale, is given by
\begin{equation}
    n_{oE}=\frac{2\pi \cdot F}{V} = n_{0}\frac{n \sigma v}{4g} \cdot
    |\overline{V_{h}}+\overline{V_{E}}|,
\end{equation}
where, with account for $V_{h} > V_{E}$, relative velocity can be
expressed as
$$
    |\overline{V_{o}}|=\sqrt{(\overline{V_{h}}+\overline{V_{E}})^{2}}=\sqrt{V_{h}^2+V_{E}^2+V_{h}V_{E}sin(\theta)} \simeq
$$
$$
\simeq V_{h}\sqrt{1+\frac{V_{E}}{V_{h}}sin(\theta)}\sim
V_{h}(1+\frac{1}{2}\frac{V_{E}}{V_{h}}sin(\theta)).
$$
Here $\theta=\omega (t-t_0)$ with $\omega = 2\pi/T$, $T=1yr$ and
$t_0$ is the phase. Then the concentration takes the form
\begin{equation}
    n_{oE}=n_{oE}^{(1)}+n_{oE}^{(2)}\cdot sin(\omega (t-t_0))
    \label{noE}
\end{equation}

So, there are two parts of the signal: constant and annual
modulation, as it is expected in the strategy of dark matter search
in DAMA experiment \cite{Bernabei:2008yi}.

Such neutral $(^4He^{++}X^{--})$ ``atoms" may provide a catalysis of
cold nuclear reactions in ordinary matter (much more effectively
than muon catalysis). This effect needs a special and thorough
investigation. On the other hand, $X^{--}$ capture by nuclei,
heavier than helium, can lead to production of anomalous isotopes,
but the arguments, presented in \cite{I,FKS,KK} indicate that their
abundance should be below the experimental upper limits.

It should be noted that the nuclear cross section of the O-helium
interaction with matter escapes the severe constraints
\cite{McGuire:2001qj} on strongly interacting dark matter particles
(SIMPs) \cite{Starkman,McGuire:2001qj} imposed by the XQC experiment
\cite{XQC}. Therefore, a special strategy of direct O-helium  search
is needed, as it was proposed in \cite{Belotsky:2006fa}.

%%%%%%%%%%%%%%%%%%%%%%%%%%%%%%%%%%%%%%%%%%%%%%%%%%%%%%%%%%%%%%%%%%%%%%%%
In underground detectors, $OHe$ ``atoms'' are slowed down to thermal
energies and give rise to energy transfer $\sim 2.5 \cdot 10^{-4}
\eV A/S_3$, far below the threshold for direct dark matter
detection. It makes this form of dark matter insensitive to the
severe CDMS constraints \cite{Akerib:2005kh}. However, $OHe$ induced
processes in the matter of underground detectors can result in observable effects.

\section{Low energy bound state of O-helium with nuclei}

In the essence, our explanation of the results of experiments DAMA/NaI and DAMA/LIBRA is based on the idea that OHe,
slowed down in the terrestrial matter and present in the matter of DAMA detectors, can form a few keV bound state with
nucleus, in which OHe is situated \textbf{beyond} the nucleus. Formation of such bound state leads to the corresponding energy release and ionization signal, detected in DAMA experiments.

\subsection{Low energy bound state of O-helium with nuclei}

We assume the following picture: at the distances larger, than its size,
OHe is neutral and it feels only Yukawa exponential tail of nuclear attraction,
due to scalar-isoscalar nuclear potential. It should be noted that scalar-isoscalar
nature of He nucleus excludes its nuclear interaction due to $\pi$ or $\rho$ meson exchange,
so that the main role in its nuclear interaction outside the nucleus plays $\sigma$ meson exchange,
on which nuclear physics data are not very definite. When the distance from the surface of nucleus becomes
smaller than the size of OHe, the mutual attraction of nucleus and OHe is changed by dipole Coulomb repulsion. Inside the nucleus strong nuclear attraction takes place. In the result the spherically symmetric potential appears,given by
\begin{equation}
U=-\frac{A_{He} A g^2 exp(-\mu r)}{r} + \frac{Z_{He} Z e^2 r_o \cdot F(r)}{r^2}.
\label{epot}
\end{equation}
Here $A_{He}=4$, $Z_{He}=2$ are atomic weight and charge of helium, $A$ and $Z$ are respectively atomic weight and charge of nucleus, $\mu$ and $g^2$ are the mass and coupling of scalar-isoscalar meson - mediator of nuclear attraction, $r_o$ is the size of OHe and $F(r)$ is its electromagnetic formfactor, which strongly suppresses the strength of dipole electromagnetic interaction outside the OHe "atom".
%Qualitatively, this potential has the form, presented on Fig. \ref{pic3}.

Schrodinger equation for this system is reduced
(taking apart the equation for the center of mass) to the equation of relative motion for the reduced mass
\begin{equation}
            m=\frac{Am_p m_o}{Am_p+m_o},
            \label{m}
 \end{equation}
where $m_p$ is the mass of proton.

In the case of orbital momentum \emph{l}=0 the wave functions depend only on \emph{r}.

\begin{figure}
    \begin{center}
        \includegraphics[width=4in]{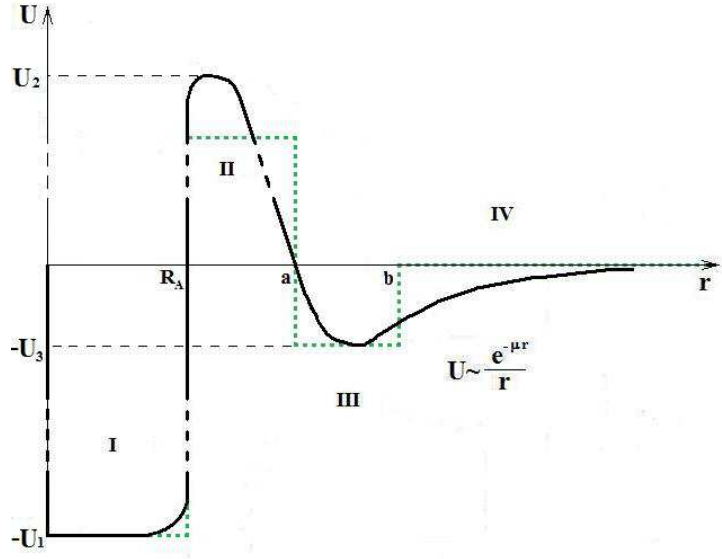}\\
        \caption{The approximation of rectangular well for potential of OHe-nucleus system.}\label{pic23}
    \end{center}
\end{figure}

To simplify the solution of Schrodinger equation we approximate the potential (\ref{epot})
by a rectangular potential that
consists of a deep potential well within the
radius of nucleus $R_A$, of a rectangular dipole Coulomb potential barrier outside its surface up to the
 radial layer $a=R_A+r_o$, where it is suppressed by the OHe atom formfactor, and of the outer potential well of the width $\sim 1/\mu$, formed by the tail of Yukawa nuclear interaction. It leads to the approximate potential, given by

\begin{equation}
    \left\{
        \begin{aligned}
        r<R_A: U=U_{1}=-\frac{4Ag^{2}exp(-\mu R_A)}{R_A},  \\
        R_A<r<a: U=U_{2}=\frac{\int_{R_A}^{R_A+r_o} \frac{2Z \alpha 4\pi(ro/x)}{x} dx}{r_o},  \\
        a<r<b: U=U_{3}=\frac{4Ag^{2}exp(-\mu (R_A+r_o) )}{R_A+r_o},  \\
        b<r: U=U_{4}=0,
        \end{aligned}
            \right.
            \label{Pot1}
 \end{equation}

    presented on Fig. \ref{pic23}.

Solutions of Schrodinger
equation for each of the four regions, indicated on Fig. \ref{pic23}, are considered in Appendix.
In the result of their sewing one obtains the condition for the existence of a low-energy level in OHe-nucleus system,
\begin{equation}
sin(k_3 b + \delta)=\sqrt{\frac{1}{2mU_3}} \cdot k_3,
\label{e21}
\end{equation}
where $k_3$ and $\delta$ are, respectively, the wave number and phase of the wave function in the region III (see Appendix for details).

\begin{figure}
    \begin{center}
        \includegraphics[width=3in]{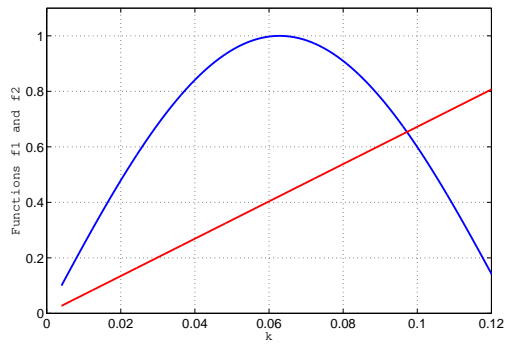}\\
        \caption{Graphical solution of transcendental equation.}\label{F12}
    \end{center}
\end{figure}

With the use of the potential (\ref{Pot1}) in the Eq.(\ref{e21}), intersection of the two lines gives graphical
solution presented on Fig. \ref{F12}.

Based on this solution one obtains from Eq.(\ref{e18})
the energy levels of a bound state in the considered potential well.

\begin{figure}
    \begin{center}
        \includegraphics[width=4in]{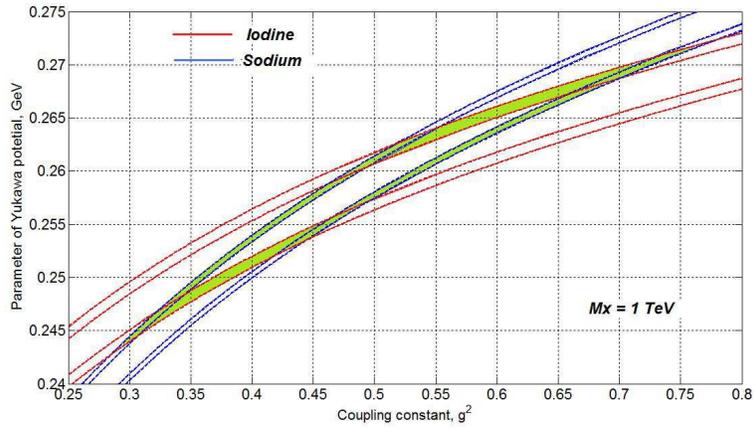}\\
        \caption{The region of parameters $\mu$ and $g^2$, for which Na and I have a level in the interval 2-6 keV. For each nucleus two narrow strips determine the region of parameters, at which the bound system of this element with OHe has a level in 2-6 keV energy range. The outer line of strip corresponds to the level of 6 keV and the internal line to the level of 2 keV. The region of intersection of strips correspond to existence of 2-6 keV levels in both OHe-Na and OHe-I systems, while the piece of strip between strips of other nucleus corresponds to the case, when OHe bound state with this nucleus has 2-6 keV level, while the binding energy of OHe with the other nuclei is less than 2 keV by absolute value.}\label{NaI}
    \end{center}
\end{figure}

The energy of this bound state and its existence strongly depend on the parameters $\mu$ and $g^2$ of nuclear potential (\ref{epot}). On the Fig. \ref{NaI} the region of these parameters, giving 2-6 keV energy level in OHe bound states with sodium and iodine are presented. In these calculations the mass of OHe was taken equal to $m_o=1 TeV$.

The rate of radiative capture of OHe by nuclei should be accurately calculated with the use of exact form of wave functions, obtained for the OHe-nucleus bound state. This work is now in progress. One can use the analogy with the radiative capture of neutron by proton, considered in textbooks (see e.g. \cite{LL4})  with the following corrections:
 \begin{itemize}
\item
  There is only E1 transition in the case of OHe capture.
\item
  The reduced masses of n-p and OHe-nucleus systems are different
\item
  The existence of dipole Coulomb barrier leads to a suppression of the cross section of OHe radiative capture.
\end{itemize}
 With the account for these effects our first estimations give the rate of OHe radiative capture, reproducing the level of signal, detected by DAMA.

Formation of OHe-nucleus bound system leads to energy release of its binding energy, detected as ionization signal in DAMA experiment. In the context of our approach the existence of annual modulations of this signal in the range 2-6 keV and absence of such effect at energies above 6 keV means that binding energy of Na-OHe and I-OHe systems should not exceed 6 keV, being in the range 2-6 keV for at least one of these elements. These conditions were taken into account for determination of nuclear parameters, at which the result of DAMA can be reproduced. At these values of $\mu$ and $g^2$ energy of OHe binding with other nuclei can strongly differ from 2-6 keV. In particular, energy release at the formation of OHe bound state with thallium can be larger than 6 keV. However, taking into account that thallium content in DAMA detector is 3 orders of magnitude smaller, than NaI, such signal is to be below the experimental errors.

It should be noted that the results of DAMA experiment exhibit also absence of annual modulations at the energy of MeV-tens MeV. Energy release in this range should take place, if OHe-nucleus system comes to the deep level inside the nucleus (in the region I of Fig. \ref{pic23}). This transition implies tunneling through dipole Coulomb barrier and is suppressed below the experimental limits.
%Preliminary results give the energy level of
%for $\mu= 320 \MeV$ and $g^2=2$, $\mu= 350 \MeV$ and $g^2=4$, $\mu= 380 \MeV$ and $g^2=10$ or for $\mu= 460 \MeV$ and $g^2=100$.

\subsection{Energy levels in other nuclei}
\begin{figure}
    \begin{center}
        \includegraphics[width=4in]{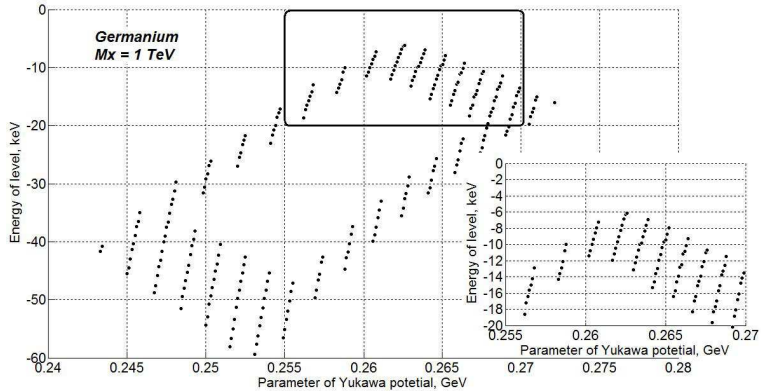}\\
        \caption{Energy levels in OHe bound system with germanium. The range of energies close to energy release in DAMA experiment is blown up to demonstrate that even in this range there is no formal intersection with DAMA results. }\label{Ge}
    \end{center}
\end{figure}

For the chosen range of nuclear parameters, reproducing the results of DAMA/NaI and DAMA/LIBRA, we can calculate the binding energy of OHe-nucleus states in nuclei, corresponding to chemical composition of set-ups in other experiments. The results of such calculation for germanium, corresponding to the detector of CDMS experiment, are presented on Fig. \ref{Ge}.
For all the parameters, reproducing results of DAMA experiment the predicted energy level of OHe-germanium bound state is beyond the range 2-6 keV, being dominantly in the range of tens - few-tens keV by absolute value. It makes elusive a possibility to test DAMA results by search for ionization signal in the same range 2-6 keV in other set-ups with content that differs from Na and I. In particular, our approach naturally predicts absence of ionization signal in the range 2-6 keV in accordance with the recent results of CDMS \cite{Kamaev:2009gp}.

\begin{figure}
    \begin{center}
        \includegraphics[width=4in]{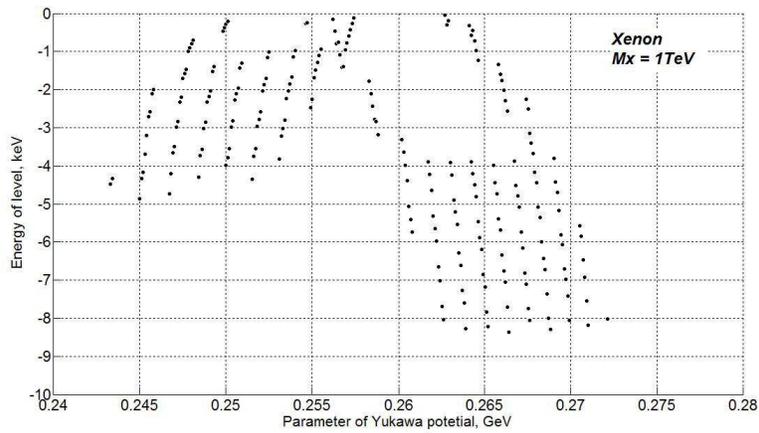}\\
        \caption{Energy levels in OHe bound system with xenon.}\label{Xe}
    \end{center}
\end{figure}

\begin{figure}
    \begin{center}
        \includegraphics[width=4in]{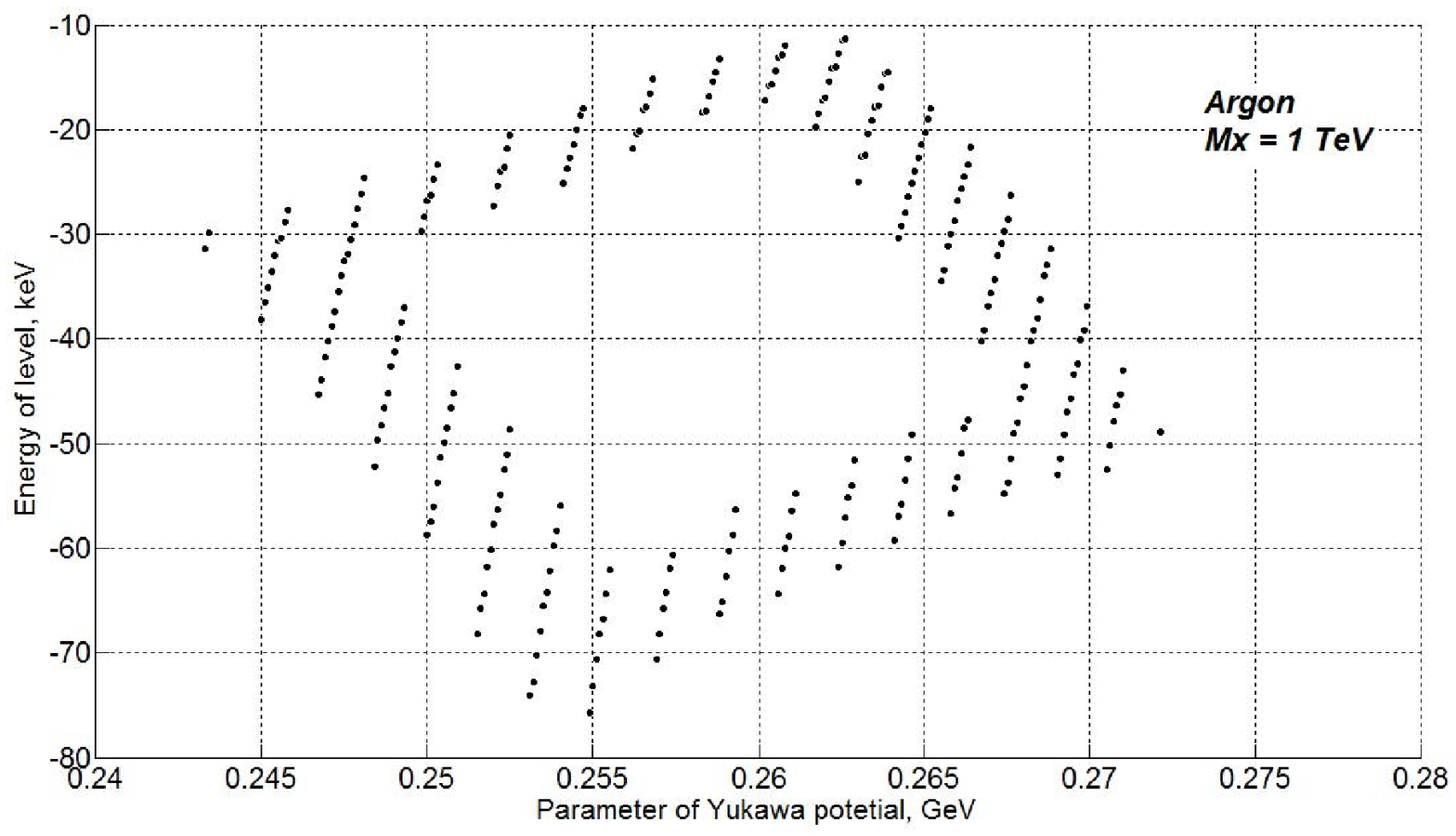}\\
        \caption{Energy levels in OHe bound system with argon.}\label{Ar}
    \end{center}
\end{figure}

\begin{figure}
    \begin{center}
        \includegraphics[width=4in]{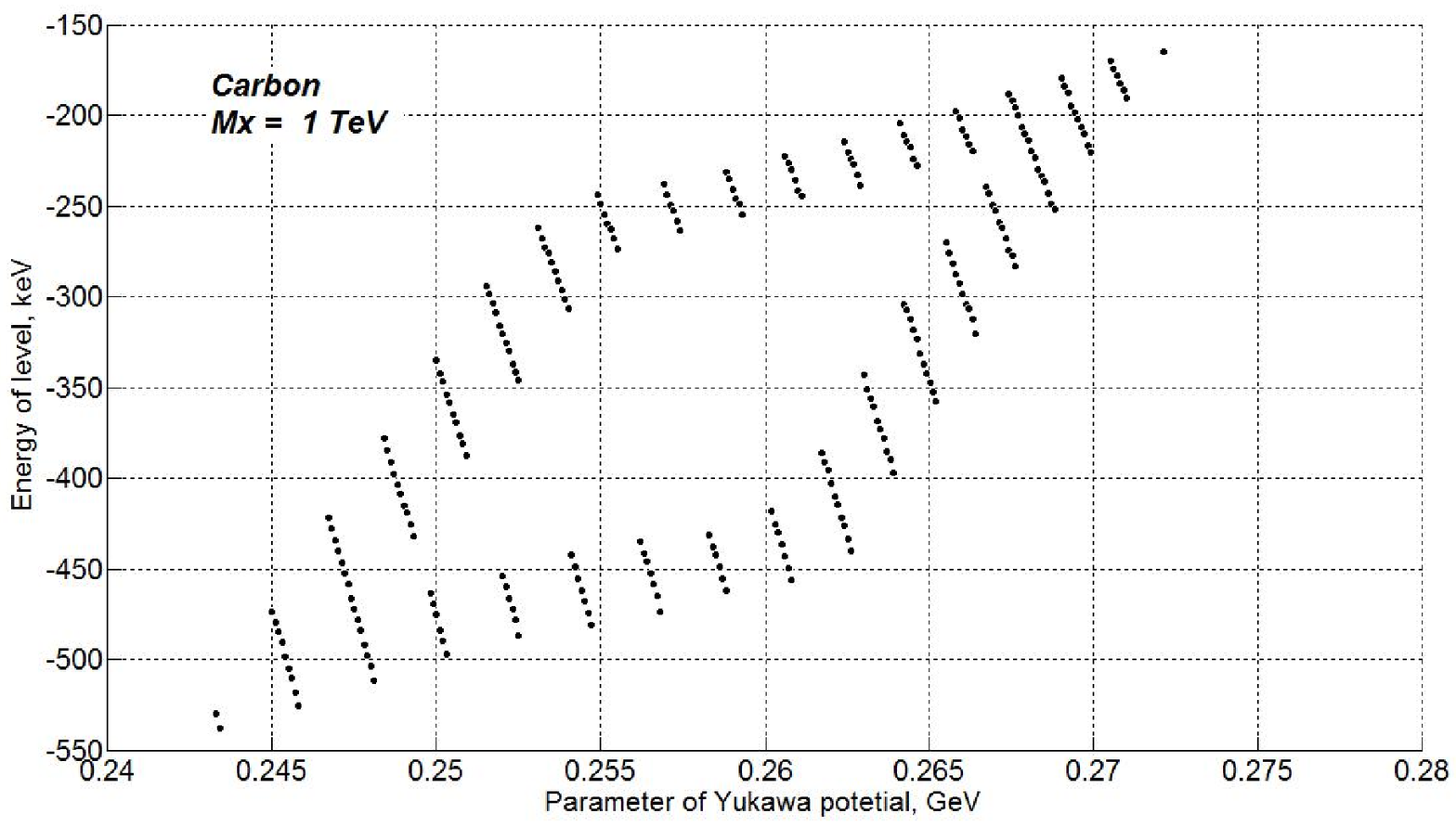}\\
        \caption{Energy levels in OHe bound system with carbon.}\label{C}
    \end{center}
\end{figure}

\begin{figure}
    \begin{center}
        \includegraphics[width=4in]{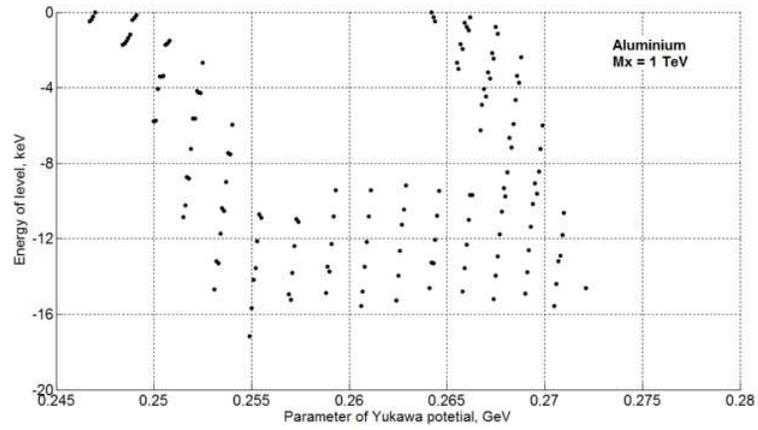}\\
        \caption{Energy levels in OHe bound system with aluminium.}\label{Al}
    \end{center}
\end{figure}

\begin{figure}
    \begin{center}
        \includegraphics[width=4in]{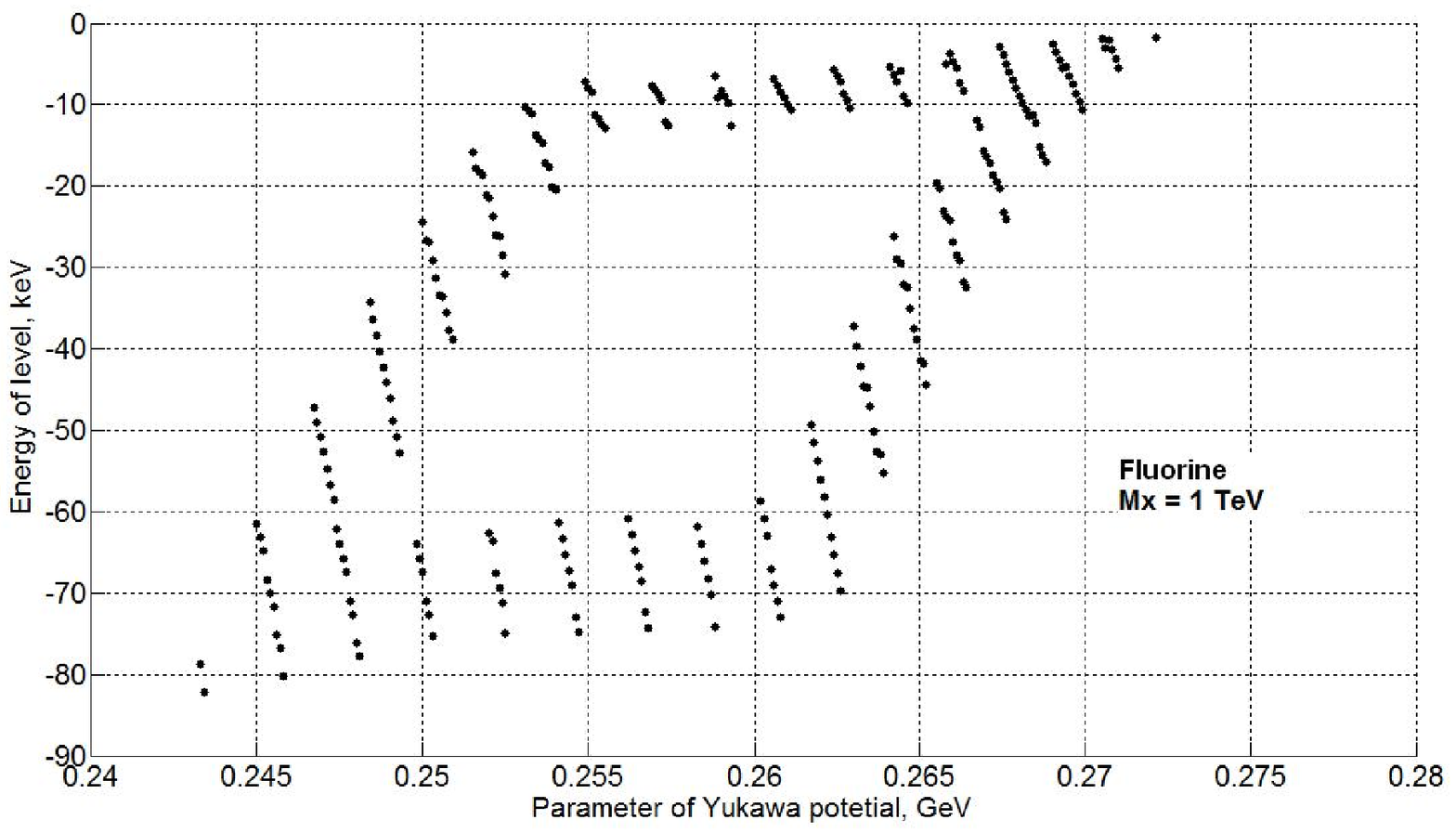}\\
        \caption{Energy levels in OHe bound system with fluorine.}\label{F}
    \end{center}
\end{figure}

\begin{figure}
    \begin{center}
        \includegraphics[width=4in]{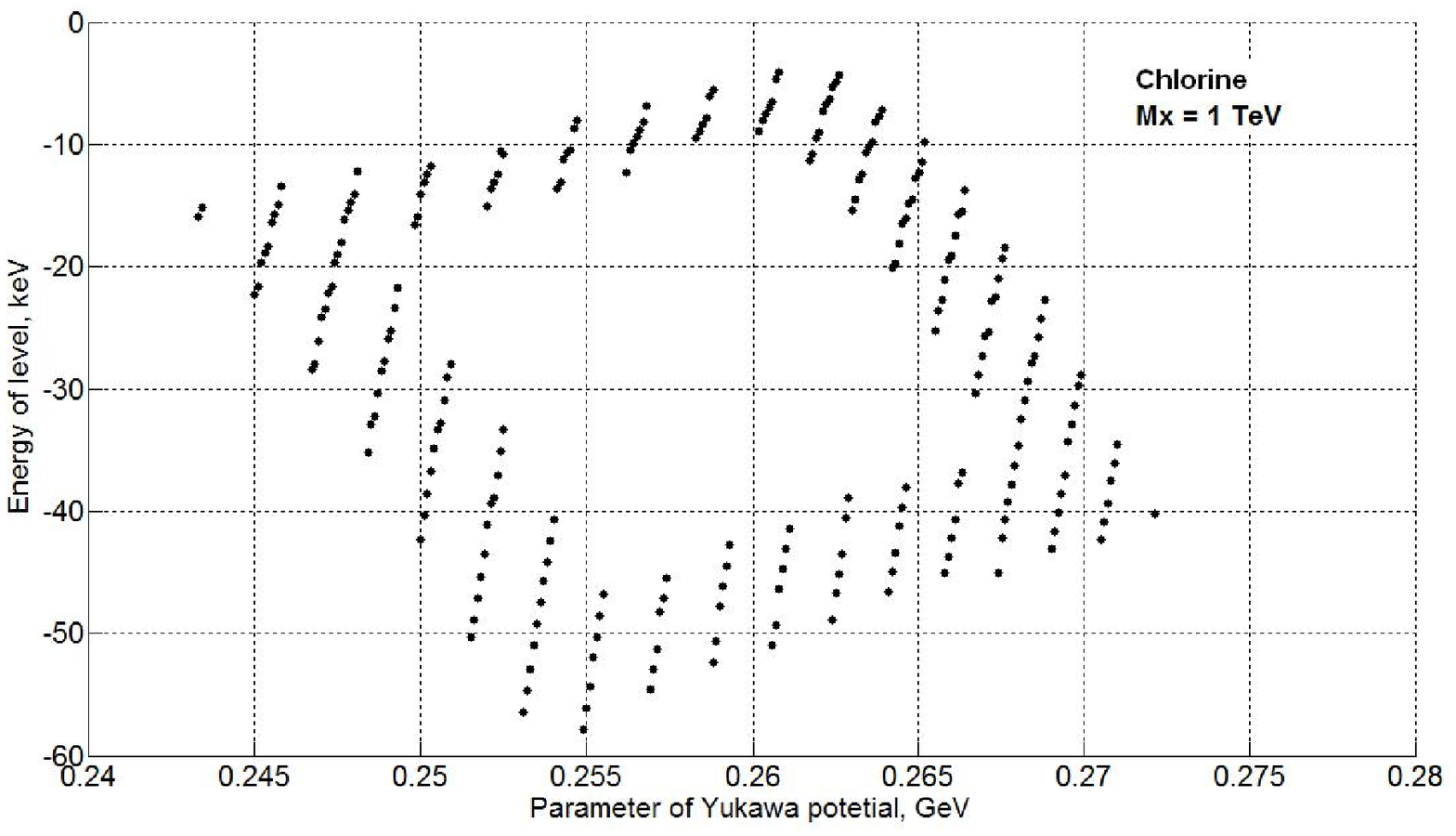}\\
        \caption{Energy levels in OHe bound system with chlorine.}\label{Cl}
    \end{center}
\end{figure}

\begin{figure}
    \begin{center}
        \includegraphics[width=4in]{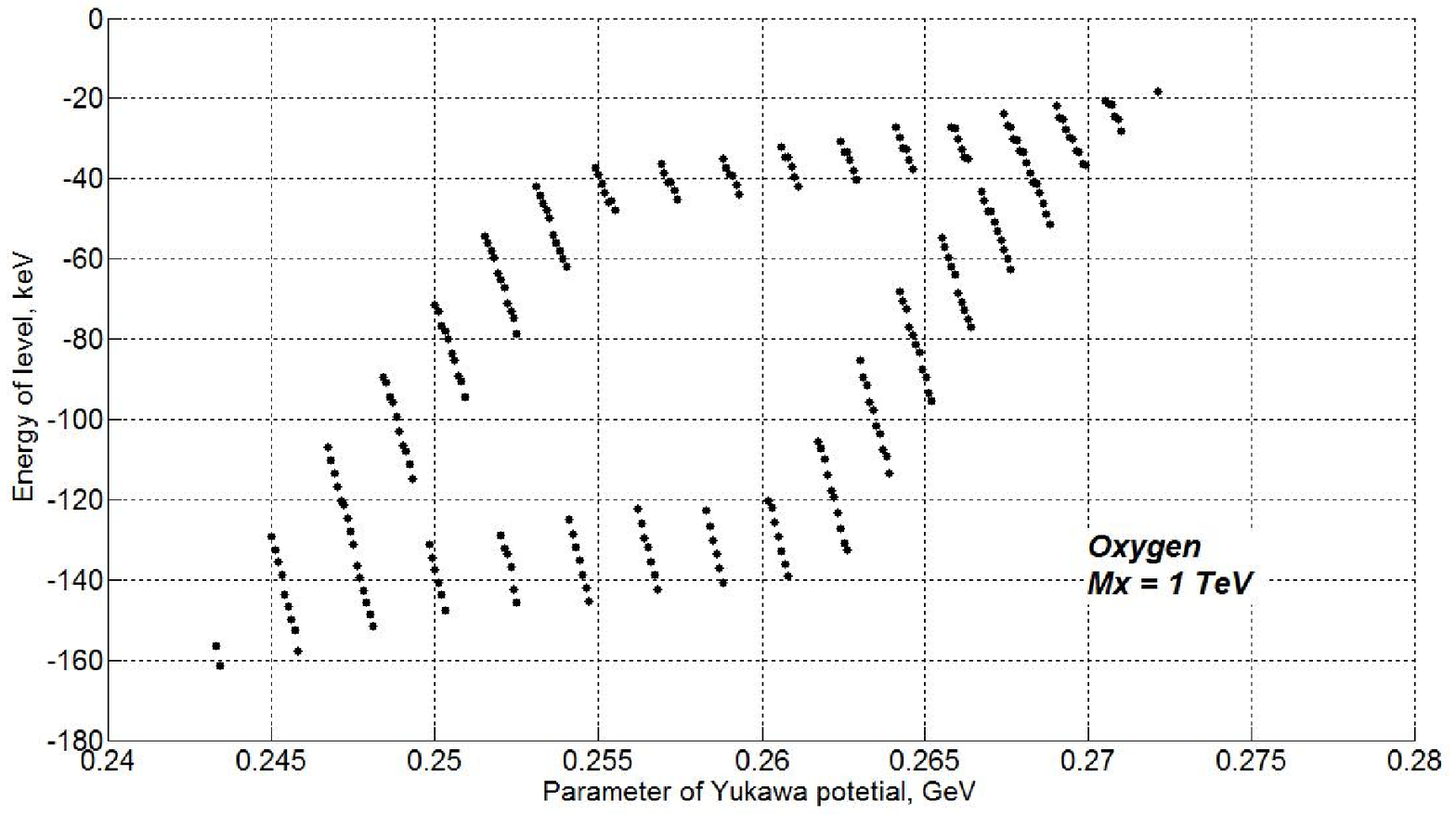}\\
        \caption{Energy levels in OHe bound system with oxygen.}\label{O}
    \end{center}
\end{figure}

We have also calculated the energies of bound states of OHe with xenon (Fig. \ref{Xe}), argon (Fig. \ref{Ar}), carbon (Fig. \ref{C}), aluminium (Fig. \ref{Al}), fluorine (Fig. \ref{F}), chlorine (Fig. \ref{Cl}) and oxygen (Fig. \ref{O}).

\subsection{Superheavy OHe}
\begin{figure}
    \begin{center}
        \includegraphics[width=4in]{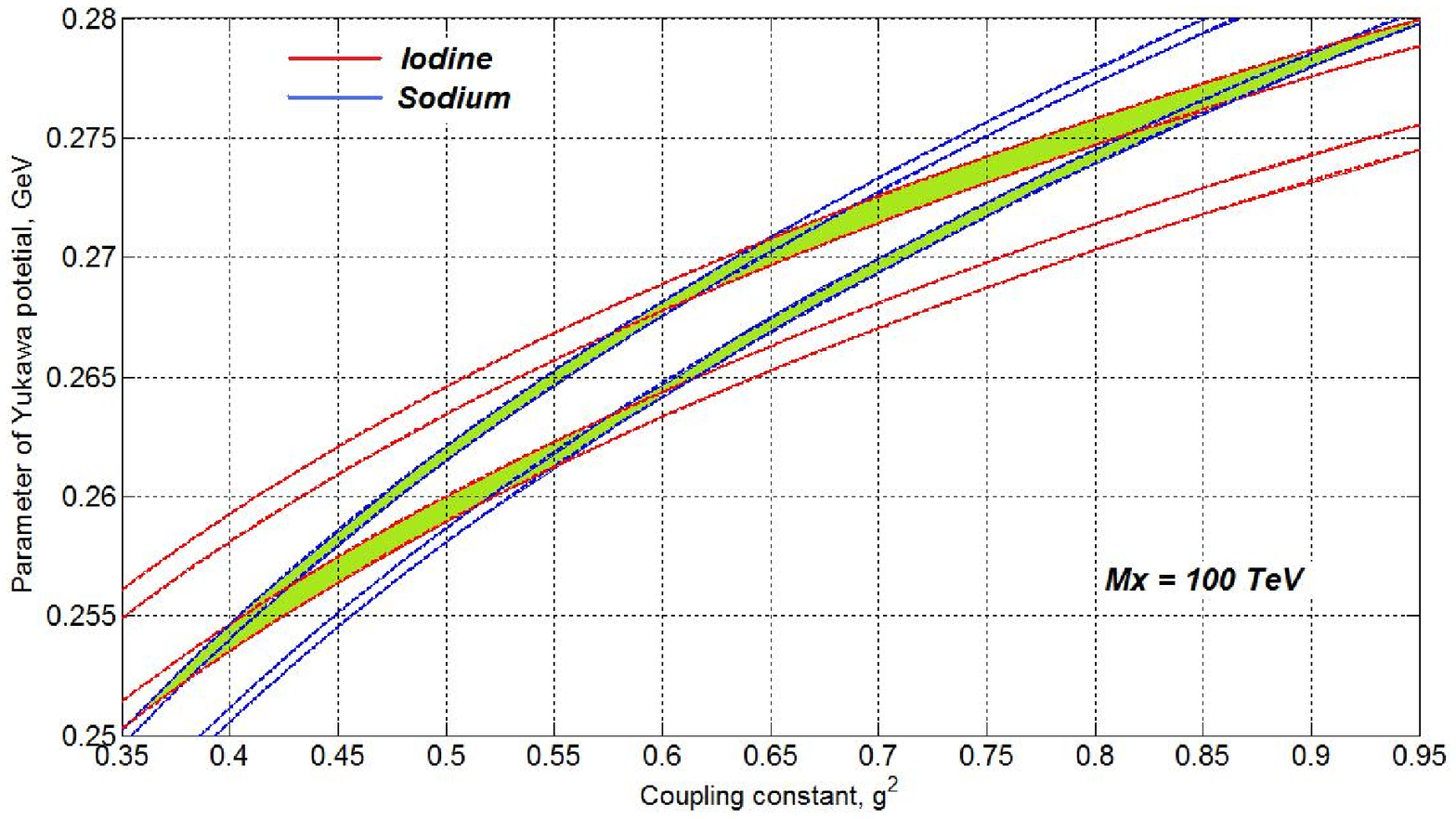}\\
        \caption{The range of parameters $\mu$ and $g^2$, for which Na and I have a level in the interval 2-6 keV for $S_3=100$. This range becomes a bit wider as compared with the case of $S_3=1$, presented on Fig. \ref{NaI}. }\label{NaI100}
    \end{center}
\end{figure}

\begin{figure}
    \begin{center}
        \includegraphics[width=4in]{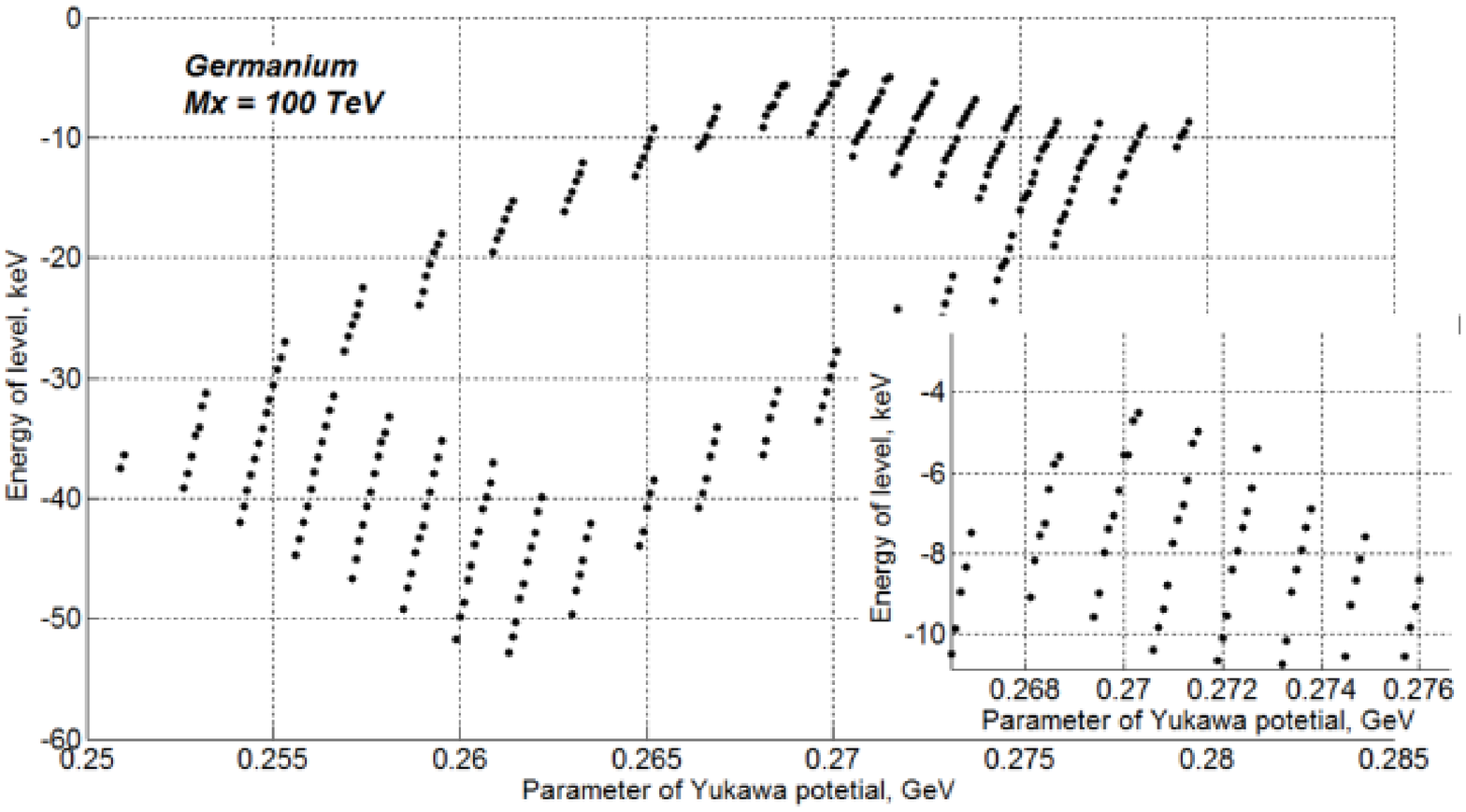}\\
        \caption{Energy levels in OHe bound system with germanium.}\label{Ge100}
    \end{center}
\end{figure}

\begin{figure}
    \begin{center}
        \includegraphics[width=4in]{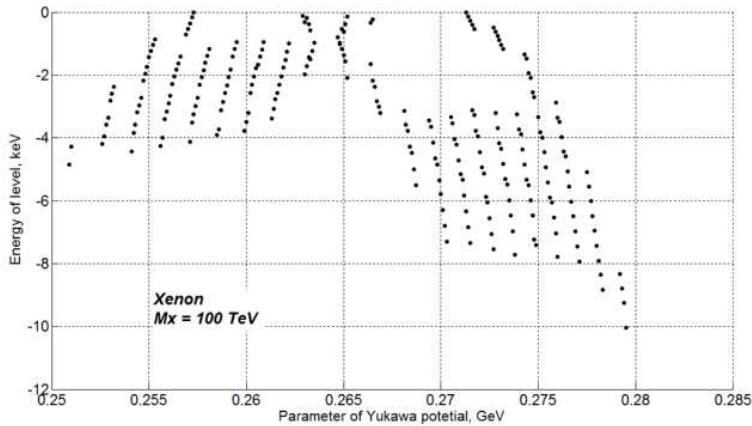}\\
        \caption{Energy levels in OHe bound system with xenon.}\label{Xe100}
    \end{center}
\end{figure}

In view of possible applications for the approach, unifying spins and charges \cite{Norma}, we consider here the case of superheavy OHe, since the candidate for $X^{--}$, coming from stable 5th generation ($\bar u_5 \bar u_5 \bar u_5$) is probably much heavier, than 1 TeV. With the growth of the mass of O-helium the reduced mass (\ref{m}) slightly grows, approaching with higher accuracy the mass of nucleus. It extends a bit the range of nuclear parameters $\mu$ and $g^2$, at which the binding energy of OHe with sodium and/or iodine is within the range 2-6 keV (see Fig. \ref{NaI100}).
At these parameters the binding energy of O-helium with germanium and xenon are presented on figures \ref{Ge100} and \ref{Xe100}, respectively. Qualitatively, these predictions are similar to the case of $S_3=1$. Though there appears a narrow window with OHe-Ge binding energy, below 6 keV for the dominant range of parameters energy release in CDMS is predicted to be of the order of few tens keV.
\section{Conclusions}

%\medskip

To conclude, the results of dark matter search in experiments
DAMA/NaI and DAMA/LIBRA can be explained in the framework of
composite dark matter scenario without contradiction with negative
results of other groups. This scenario can be realized in different
frameworks, in particular in Minimal Walking Technicolor model or in
the approach unifying spin and charges and contains distinct
features, by which the present explanation can be distinguished from
other recent approaches to this problem \cite{Edward} (see also
review and more references in \cite{Gelmini}).

Our explanation is based on the mechanism of low energy binding of OHe with nuclei.
We have found that within the uncertainty of nuclear physics parameters there exists a range at which OHe
binding energy with sodium and/or iodine is in the interval 2-6 keV. Radiative capture of OHe to this bound state leads to the corresponding energy release observed as an ionization signal
in DAMA detector.

OHe concentration in the matter of underground detectors is determined by the equilibrium between the incoming cosmic flux of OHe and diffusion towards the center of Earth. It is rapidly adjusted and follows the
change in this flux with the relaxation time of few
minutes. Therefore the rate of radiative capture of OHe should experience annual modulations reflected in annual modulations of the ionization signal from these reactions.

%The method to calculate the rate of OHe reactions was developed and
%the calculated total amount of such events is shown to be consistent
%with the results of DAMA/NaI and DAMA/LIBRA experiments for the mass
%of OHe around 1 TeV. This method can be applied to the analysis of
%the whole set of inelastic processes, induced by O-helium in matter.

An inevitable consequence of the proposed explanation is appearance
in the matter of DAMA/NaI or DAMA/LIBRA detector anomalous
superheavy isotopes of sodium and/or iodine,
having the mass roughly by $m_o$ larger, than ordinary isotopes of
these elements. If the atoms of these anomalous isotopes are not
completely ionized, their mobility is determined by atomic cross
sections and becomes about 9 orders of magnitude smaller, than for
O-helium. It provides their conservation in the matter of detector. Therefore mass-spectroscopic
analysis of this matter can provide additional test for the O-helium
nature of DAMA signal. Methods of such analysis should take into account
the fragile nature of OHe-Na (and/or OHe-I) bound states. Their binding energy is only few keV.

With the account for high sensitivity of our results to the values of uncertain nuclear parameters
and for the approximations, made in our calculations, the presented results can be considered
only as an illustration of the possibility to explain puzzles of dark matter search in
the framework of composite dark matter scenario. However, even at the present level of
our studies we can make a conclusion that the ionization signal expected in detectors
with the content, different from NaI, can be dominantly in the energy range beyond 2-6 keV.
Therefore test of results of DAMA/NaI and DAMA/LIBRA experiments by other experimental groups can become a very nontrivial task.

%\bigskip

%{\centering{ \large \textbf{Acknowledgments}} }

\section {Acknowledgments}

%\medskip

We would like to thank Jean Pierre Gazeau for discussions.

\section*{Appendix. Solution of Schrodinger equation for rectangular well}
In the 4 regions, indicated on Fig. \ref{pic23}, Schrodinger
equation has the form
\begin{equation}
I: \frac{1}{r}\frac{d^2}{dr^2}(r\psi_1)+k_1(r)^2\psi_1=0,
k_1(r)=k_1=\sqrt{2m(U_1-|E|)};
\end{equation}

\begin{equation}
II: \frac{1}{r}\frac{d^2}{dr^2}(r\psi_2)+k_2(r)^2\psi_2=0,
k_2(r)=k_2=\sqrt{2m(U_2-|E|)};
\end{equation}

\begin{equation}
III: \frac{1}{r}\frac{d^2}{dr^2}(r\psi_3)+k_3(r)^2\psi_3=0,
k_3(r)=k_3=\sqrt{2m(U_3-|E|)};
\end{equation}

\begin{equation}
IV: \frac{1}{r}\frac{d^2}{dr^2}(r\psi_4)-k_4(r)^2\psi_4=0,
k_4(r)=k_4=\sqrt{2m|E|}.
\end{equation}

The wave functions in these regions with the account for the boundary conditions have the form \cite{LL3}

\begin{equation}
I: \psi_1=A\frac{\sin (k_1 r)}{r};
\end{equation}

\begin{equation}
II: \psi_2=\frac{B_1\cdot exp(-k_2 r)+B_2\cdot exp(k_2 r)}{r};
\end{equation}

\begin{equation}
III: \psi_3=C\frac{\sin (k_3 r + \delta)}{r}
\end{equation}

\begin{equation}
IV: \psi_4=D\frac{\exp (-k_4 r)}{r}
\end{equation}

The conditions of continuity of a logarithmic derivative
$\frac{\psi_i'}{\psi_i}=\frac{\psi_{i+1}'}{\psi_{i+1}}$ от
$\emph{r}\psi$ at the boundaries of these regions \emph{$r=R_A$},
\emph{$r=a$} and \emph{$r=b$} are given by

\begin{equation}
I - II: k_1\cdot ctg(k_1 R_A) = k_2\cdot \frac{exp(k_2 R_A)-F\cdot
exp(-k_2 R_A)}{exp(k_2 R_A)+F\cdot exp(-k_2 R_A)},
\label{e12}
\end{equation}

\begin{equation}
II - III: k_3\cdot ctg(k_3 a + \delta) = k_2\cdot \frac{exp(k_2
a)-F\cdot exp(-k_2 a)}{exp(k_2 a)+F\cdot exp(-k_2 a)},
\label{e13}
\end{equation}

\begin{equation}
III - IV: k_3\cdot ctg(k_3 b + \delta) = -k_4,
\label{e14}
\end{equation}

where
\begin{equation}
F = B_1/B_2.
\end{equation}

Now we can solve this system of equations for 3 variables. It follows from Eq. (\ref{e12}) that
\begin{equation}
F = exp(2 k_2 R_A) \cdot \frac{k_2-k_1\cdot ctg(k_1
R_A)}{k_2+k_1\cdot ctg(k_1 R_A)},
\end{equation}

and from Eq. (\ref{e13})

\begin{equation}
\delta=-k_3 a + arcctg(\frac{k_2}{k_3}\cdot \frac{exp(k_2 a)-F\cdot
exp(-k_2 a)}{exp(k_2 a)+F\cdot exp(-k_2 a)}).
\end{equation}

Since

\begin{equation}
E=U(r)-\frac{k^2}{2m},
\label{e18}
\end{equation}

one has

\begin{equation}
k_4=\sqrt{2mU-k_3^{2}},
\end{equation}

Then Eq.(\ref{e14}) has the form
\begin{equation}
k_3^{2}[\frac{1}{sin^{2}(k_3 b + \delta)}-1]=2mU_3-k_3^{2},
\end{equation}

%\bigskip

%{\centering{ \large \textbf{References}} }

%\section*{References}

%\medskip

\end{document}